\documentclass[10pt,conference,twocolumn]{IEEEtran}

\usepackage{amsmath}
\usepackage{amssymb}
\usepackage{latexsym}
\usepackage{amsfonts}

\usepackage{LATEXCAD}

\begin{document}
\pagestyle{headings}  
\vspace{1cm}

\title{On the  $k$-error linear complexity for $2^n$-periodic sequences via Cube Theory }

\author{
\authorblockN{Jianqin Zhou}
\authorblockA{ 1. Department of Computing, Curtin University, Perth, WA 6102 Australia\\
2. Computer Science School, Anhui Univ. of
Technology, Ma'anshan, 243002 China\\ \ \ zhou9@yahoo.com\\
\ \\
Wanquan Liu\\
Department of Computing, Curtin University, Perth, WA 6102 Australia\\
 w.liu@curtin.edu.au
 }
}
\maketitle              

\begin{abstract}
The linear complexity and k-error linear complexity of a sequence
have been used as important measures of keystream strength, hence
designing a sequence with high linear complexity and $k$-error
linear complexity is a popular research topic in cryptography.  In
this paper, the concept of stable $k$-error linear complexity is
proposed to study sequences with stable and large $k$-error linear
complexity. In order to study  k-error linear complexity of binary
sequences with period $2^n$, a new tool called cube theory is
developed. By using the cube theory, one can easily construct
sequences with the maximum stable $k$-error linear complexity. For
such purpose, we first prove that a binary sequence with period
$2^n$ can be decomposed into some disjoint cubes and further give a
general decomposition approach. Second, it is proved that the
maximum  $k$-error linear complexity is $2^n-(2^l-1)$ over all
$2^n$-periodic binary sequences, where $2^{l-1}\le k<2^{l}$.
Thirdly, a characterization is presented about
 the $t$th ($t>1$) decrease  in the
$k$-error linear complexity for a $2^n$-periodic binary sequence $s$
and this is a continuation of Kurosawa et al. recent work for the
first decrease of k-error linear complexity. Finally, A counting
formula for $m$-cubes with the same linear complexity is derived,
which is equivalent to the counting formula for $k$-error vectors.
The counting formula
of $2^n$-periodic binary sequences which can be decomposed
into more than one cube is also investigated, which extends an important result by Etzion et al..


\end{abstract}

\section{Introduction}


It is well known that stream ciphers have broad applications in
network security. The linear complexity of a sequence $s$, denoted
as $L(s)$, is defined as the length of the shortest linear feedback
shift register (LFSR) that can generate $s$. The concept of linear
complexity is very useful in the study of the security for stream
ciphers. A necessary condition for the security of a key stream
generator is that it produces a sequence with high linear
complexity. However, high linear complexity can not necessarily
guarantee the sequence is secure. The linear complexity of some
sequences is unstable. If a small number of changes to a sequence
greatly reduce its linear complexity, then the resulting key stream
would be cryptographically weak. Ding, Xiao and Shan in their book
\cite{Ding} noticed this problem first, and presented the concepts
of weight complexity and sphere complexity. Stamp and Martin
\cite{Stamp} introduced $k$-error linear complexity, which is
similar to the sphere complexity, and proposed the concept of
$k$-error linear complexity profile. Suppose that $s$ is a sequence
over $GF(q)$ with period $N$. For $k(0\le k\le N)$, the $k$-error
linear complexity of $s$, denoted as $L_k(s)$, is defined as the
smallest linear complexity that can be obtained when any $k$ or
fewer of the terms of the sequence are changed within one period.
For small $k$, Niederreiter \cite{Niederreiter} presented some
sequences over $GF(q)$ which possess large linear complexity and
$k$-error linear complexity. By using the generalized discrete
Fourier transform, Hu and Feng \cite{Hu} constructed some periodic
sequences over $GF(q)$ which possess very large 1-error linear
complexity.

One important result, proved by Kurosawa et al. in \cite{Kurosawa}
is that the minimum number $k$ for which the $k$-error linear
complexity of a $2^n$-periodic binary sequence $s$ is strictly less
than the linear complexity $L(s)$ of $s$ is determined by
$k_{\min}=2^{W_H(2^n-L(s))}$, where $W_H(a)$ denotes the Hamming
weight of the binary representation of an integer $a$. In
\cite{Meidl}, for the period length $p^n$, where $p$ is an odd prime
and 2 is a primitive root modulo $p^2$, the  relationship is showed
between the linear complexity and the minimum value $k$ for which
the $k$-error linear complexity is strictly less than the linear
complexity. In \cite{Zhou}, for sequences over $GF(q)$ with period
$2p^n$, where $p$ and $q$ are odd primes, and $q$ is a primitive
root modulo $p^2$, the minimum value $k$ is presented for which the
$k$-error linear complexity is strictly less than the linear
complexity. For $k=1,2$, Meidl \cite{Meidl2005} characterized the
complete counting functions on the $k$-error linear complexity of
$2^n$-periodic binary sequences with the maximal possible linear
complexity $2^n$.
 Fu et al. \cite{Fu2006} studied the linear complexity and
the 1-error linear complexity of $2^n$-periodic binary sequences,
and then  to characterize such sequences with fixed 1-error linear
complexity. For $k=2,3$, Zhu and Qi \cite{Zhu} further derived the
complete counting functions on the $k$-error linear complexity of
$2^n$-periodic binary sequences with linear complexity $2^n-1$. The
complete counting functions for the number of $2^n$-periodic binary
sequences with  3-error linear complexity  are given by Zhou and Liu
recently in \cite{Zhou_Liu}.

The motivation of studying the stability of linear complexity is
that changing a small number of elements in a sequence may lead to a
sharp decline of its linear complexity. Therefore we really need to
study such stable sequences in which even a small number of changes
do not reduce their linear complexity. The stable $k$-error linear
complexity is introduced first in this paper to deal with this
problem. Suppose that $s$ is a sequence over $GF(q)$ with period
$N$. For $k(0\le k\le N)$, the $k$-error linear complexity of $s$ is
defined as stable when any $k$ or fewer of the terms of the sequence
are changed within one period, the linear complexity does not
decline. In this case, the $k$-error linear complexity of sequence
$s$ is equivalent to its linear complexity.
The concept of stable k-error linear complexity is very
important  and we will investigate
it in this paper.

Algebra \cite{Meidl,Meidl2005,Fu2006,Zhu} and discrete Fourier
transform \cite{Hu} are two important tools to study  the $k$-error
linear complexity for periodic sequences. Etzion et al. \cite{Etzion}
studied the sequences using algebra with two $k$-error linear
complexity values exactly, namely its $k$-error linear complexity is
only $L(s)$ or 0. To further investigate the sequences in general
case, we develop a new tool called cube theory to study the stable
$k$-error linear complexity of binary sequences with period $2^n$.
By using the proposed cube theory, we are capable of investigating
the $k$-error linear complexity for periodic sequences from a new
perspective.
 One significant benefit is to construct sequences with
the maximum stable $k$-error linear complexity. Some examples are
also given to illustrate the approach. Furthermore, it is proved
that a binary sequence with period $2^n$ can be decomposed into some
disjoint cubes and  we give a general decomposition approach, which
is called a standard cube decomposition in this paper. With such
decomposition, it is proved that the maximum $k$-error linear
complexity is $2^n-(2^l-1)$ over all $2^n$-periodic binary
sequences, where $2^{l-1}\le k<2^{l}$.
 Kurosawa et al. in \cite{Kurosawa} studied the minimum number $k$ for which the first decrease occurs for the  $k$-error linear complexity. With the cube theory, we further
 characterize the minimum number $k$ for which
the $t$th decrease occurs in the $k$-error linear complexity, $t>1$.

From the perspective of cube theory proposed, we can easily perceive the  core problem and difficulty points of the  $k$-error linear
complexity for a  $2^n$-periodic binary sequence with more
than one cube.

Technically, for $2^n$-periodic binary sequences $s$ and $e$, if
$W_H(e)=k_{\min}$ and $L(s+e)<L(s)$, then we define the sequence $e$
as a $k$-error vector associated with $s$. A $k$-error vector is in
fact an $m$-cube with the same linear complexity $L(s)$ as shown in
this paper. Based on this observation, the counting formula of
$m$-cubes with the same linear complexity will be derived with an
approach much different from that used in  \cite{Etzion} by Etzion
et al.. Based on the independence among cubes of a sequence, we propose to construct each cube independently.
As a consequence, the counting formula of $2^n$-periodic
binary sequences which can be decomposed into more than one cube is
also investigated.

The rest of this paper is organized as follows. In  Section II, some
preliminary results are presented.
 In Section III, the definition of  cube theory and our main results
are reported. Our conclusion is given in Section IV.

\section{Preliminaries}

We will consider sequences over $GF(q)$, which is the finite field
of order $q$. Let $x=(x_1,x_2,\cdots,x_n)$ and
$y=(y_1,y_2,\cdots,y_n)$ be vectors over $GF(q)$. Then we define
$$x+y=(x_1+y_1,x_2+y_2,\cdots,x_n+y_n).$$
If $q=2$, we denote $x+y$ as $x\bigoplus y$ as well.

When $n=2m$, we define
$Left(x)=(x_1,x_2,\cdots,x_m)$ and $Right(x)=(x_{m+1},x_{m+2},\cdots,x_{2m})$.

The Hamming weight of an $N$-periodic sequence $s$ is defined as the
number of nonzero elements in per period of $s$, denoted by
$W_H(s)$. Let $s^N$ be one period of $s$. If $N=2^n$, $s^N$ is also
denoted as $s^{(n)}$. The distance of two elements is defined as the
difference of their indexes. Specifically, for  an $N$-periodic sequence $s=\{s_0, s_1, s_2, s_3,
\cdots, \}$, the distance of $s_i,s_j$ is $j-i$, where $0\le i\le j\le N$.

The generating function of a sequence $s=\{s_0, s_1, s_2, s_3,
\cdots, \}$  is defined by $$s(x)=s_0+ s_1x+ s_2x^2+ s_3x^3+
\cdots=\sum\limits^\infty_{i=0}s_ix^i$$

The generating function of a finite sequence $s^N=\{s_0, s_1, s_2,
 \cdots, s_{N-1},\}$ is defined by $s^N(x)=s_0+ s_1x+ s_2x^2+
\cdots+s_{N-1}x^{N-1}$. If  $s$  is a periodic sequence with the first
period $s^N$, then,
\begin{eqnarray}
s(x) &=& s^N(x)(1+ x^N+ x^{2N}+ \cdots)=\frac{s^N(x)}{1-x^N}\notag\\
&=&\frac{s^N(x)/\gcd(s^N(x),1-x^N)}{(1-x^N)/\gcd(s^N(x),1-x^N)}\notag\\
&=&\frac{g(x)}{f_s(x)}\label{formula01}
\end{eqnarray}
where $f_s(x)=(1-x^N)/\gcd(s^N(x),1-x^N),
g(x)=s^N(x)/\gcd(s^N(x),1-x^N)$.

Obviously, one has $\gcd(g(x),f_s(x))=1,$ and $
\deg(g(x))<\deg(f_s(x))$. $f_s(x)$ is called  the minimal polynomial
of  $s$ , and the degree of $f_s(x)$ is called the linear complexity
of  $s$ , that is $\deg(f_s(x))=L(s)$.

Suppose that $N=2^n$, then $1-x^N=1-x^{2^n}=(1-x)^{2^n}=(1-x)^N$.
Thus for binary sequences with period $2^n$, to find its linear
complexity is equivalent to computing
 the degree of factor $(1-x)$ in $s^N(x)$.

The linear complexity of a $2^n$-periodic binary sequence $s$
  can be recursively computed by the Games-Chan algorithm \cite{Games} as follows.

\noindent {\bf Algorithm  2.1}

\noindent {\bf Input}: A $2^n$-periodic binary sequence $s=[Left(s),Right(s)]$, $c=0$.

\noindent {\bf Output}:  $L(s)=c$.

\noindent Step 1. If $Left(s)=Right(s)$, then deal with $Left(s)$ recursively. Namely, $L(s)=L(Left(s))$.

\noindent Step 2. If $Left(s)\neq Right(s)$, then $c=c+2^{n-1}$ and deal with $Left(s)\bigoplus Right(s)$ recursively.
Namely, $L(s)=2^{n-1}+L(Left(s)\bigoplus Right(s))$.

\noindent Step 3. If $s=(a)$, then if $a=1$ then $c=c+1$.

\

For $k(0\le k\le N)$, the $k$-error linear complexity of $s$ is
defined as stable when any $k$ or fewer of the terms of the sequence
are changed within one period, the linear complexity does not
decline. In this case, the $k$-error linear complexity of sequence
$s$ is equivalent to its linear complexity. The following three
lemmas are  well known results on $2^n$-periodic binary sequences
and required in this paper. Please refer to
\cite{Meidl2005,Fu2006,Zhu,Zhou_Liu} for details.

\noindent {\bf Lemma  2.1} Suppose that $s$ is a binary sequence
with period $N=2^n$, then $L(s)=N$ if and only if the Hamming weight
of a period of the sequence is odd.

If an element $1$ is removed from a sequence whose Hamming weight is
odd, the Hamming weight of the sequence will be changed to even, so
the main concern hereinafter is about sequences whose Hamming
weights are even.

\noindent {\bf Lemma 2.2}  Let $s_1$ and $s_2$ be binary sequences
with period $N=2^n$. If $L(s_1)\ne L(s_2)$, then
$L(s_1+s_2)=\max\{L(s_1),L(s_2)\} $; otherwise if $L(s_1)= L(s_2)$,
then $L(s_1+s_2)<L(s_1)$.

Suppose that the linear complexity of $s$ can decrease when at least
$k$ elements of $s$ are changed. By Lemma 2.2, the linear complexity
of the binary sequence, in which elements at exactly those $k$
positions are all nonzero, must be $L(s)$. Therefore, for the
computation of $k$-error linear complexity, we only need to find the
binary sequence whose Hamming weight is minimum and its linear
complexity is $L(s)$.

\noindent {\bf Lemma  2.3} Suppose that $E_i$ is a $2^n$-periodic
binary sequence  with one nonzero element at position $i$ and 0
elsewhere in each period, $0\le i\le N$. If $j-i=2^r(1+2a), a\ge0,
0\le i<j<N, r\ge0$, then $L(E_i +E_j)=2^n-2^r$.

Denote $E_{ij}$ as a binary sequence with period $2^n$, and it has
only 2 nonzero elements in a period. If there are only 2 adjacent
positions with nonzero elements  in $E_{ij}$, then its linear
complexity is $2^n-1$, namely $E_{ij}$ is a sequence with even
Hamming weight and the largest linear complexity. According to Lemma
2.2, if sequence $s$ can be decomposed into the superposition of
several $E_{ij}$, in which each has linear complexity  $2^n-1$, and
the number of such $E_{ij}$ is odd, then $L(s) =2^n-1$. After a
symbol of $s$ is changed, its Hamming weight will be odd, so its
linear complexity will be $2^n$, namely the 1-error linear
complexity of sequence $s$ is $2^n-1$. So we have the following
result.

\noindent {\bf Proposition  2.1}  If $s$ is a binary sequence with period
$2^n$, then its maximum 1-error linear complexity is $2^n-1$.

In order to discuss the maximal 2-error linear complexity of a
binary sequence with period $2^n$, we now consider a binary sequence
which has only 4 positions with nonzero elements.  Please refer to
\cite{Zhou_Liu} for the proof of Lemma  2.4.

\noindent {\bf Lemma  2.4} If $s$ is a binary sequence with period
$N=2^n$ and there are only four non-zero elements, thus $s$ can be
decomposed into the superposition of  $E_{ij}$ and  $E_{kl}$.
Suppose that non-zero positions of  $E_{ij}$ are $i$ and $j$,
$j-i=2^d$(1+2u), and non-zero positions of  $E_{kl}$ are $k$ and
$l$, $l-k=2^e$(1+2v), $i<k, k-i=2c+1$. If $d=e$, the linear
complexity is $2^n-(2^d+1)$, otherwise the linear complexity is
$2^n-2^{\min(d,e)}$.

More specifically, if we put the requirement on $E_{ij}$ and
$E_{kl}$, we will have the following result.

\noindent {\bf Lemma  2.5} If $s$ is a binary sequence with period
$2^n$ and there are only 4 non-zero elements, and $s$ can be
decomposed into the superposition of $E_{ij}$ and  $E_{kl}$, in which each has linear
complexity $2^n-1$, then the linear complexity of $s$ is $2^n-(2^d+1)$
or $2^n-2^d, d>0$.

\begin{proof}\
 Suppose that the non-zero positions of $E_{ij}$ are $i$ and $j$ with linear complexity being $2^n-1$, and $j-i=2a+1$,
 and non-zero positions of $E_{kl}$ are $k$ and $l$, whose linear complexity is also $2^n-1$,with  $i<k, l-k=2b+1$.

Next we will investigate the problem with the following 6 cases:

 1) $i<k<l<j$, and $k-i=2c$.

As $j-i=2a+1,l-k=2b+1$, so $$j-l=2a+1-(2b+1+2c)=2(a-b-c)$$

If $j-l=2^d+2u2^d, k-i=2^e+2v2^e$, without loss of generality,
assume $d<e$, by Lemma 2.2, $L(s)= 2^n-2^d$, $d>0$.

 If $d=e$, by Lemma 2.4, since
$l-i=2(b+c)+1$, so $L(s)= 2^n-(2^d+1)$.

2) $i<k<l<j$, and $k-i=2c+1$.

As $j-i=2a+1,l-k=2b+1$, so $l-i=2b+1+2c+1=2(b+c+1),
j-k=2a+1-(2c+1)=2(a-c)$

 If $j-k=2^d+2u2^d, l-i=2^e+2v2^e,$ without loss of
generality, assume $d<e$, by Lemma 2.2, $L(s) = 2^n-2^d$, $d>0$.

Since $k-i=2c+1$, by Lemma 2.4, if $d=e$, then $L(s) = 2^n-(2^d+1)$.

3) $i<k<j<l$, and $k-i=2c$.

As $j-i=2a+1,l-k=2b+1$,so $j-k=2a+1-2c=2(a-c)+1,
l-j=2b+1-[2(a-c)+1]=2(b+c-a)$

If $l-j=2^d+2u2^d, k-i=2^e+2v2^e$, without loss of generality,
assume $d<e$, by Lemma 2.2, $L(s) = 2^n-2^d,d>0$.

Since $j-i=2a+1$, by Lemma 2.4, if $d=e$, then $L(s) = 2^n-(2^d+1)$.

4) $i<k<j<l$, and $k-i=2c+1$.

As $j-i=2a+1,l-k=2b+1$,so $j-k=2a+1-(2c+1)=2(a-c),
l-i=2b+1+2c+1=2(b+c+1)$.

 If $l-i=2^d+2u2^d, j-k=2^e+2v2^e$, without loss
of generality, assume $d<e$, by Lemma 2.2,$L(s) = 2^n-2^d$,$d>0$.

Since $k-i=2c+1$, by Lemma 2.4, if $d=e$, then $L(s) = 2^n-(2^d+1)$.

5) $i<j<k<l$, and $k-i=2c$.

As $j-i=2a+1,l-k=2b+1$, so $k-j=2c-(2a+1)=2(c-a)-1,
l-j=2b+1+[2(c-a)-1]=2(b+c-a)$

 If $l-j=2^d+2u2^d, k-i=2^e+2v2^e$, without
loss of generality, assume $d<e$, by Lemma 2.2, $L(s) = 2^n-2^d,d>0$.

Note that $j-i=2a+1$, by Lemma 2.4, if $d=e$, then $L(s)=2^n-(2^d+1)$.

6) $i<j<k<l$, and $k-i=2c+1$.

As $j-i=2a+1,l-k=2b+1$, so $k-j=2c+1-(2a+1)=2(c-a),
l-i=2b+1+2c+1=2(b+c+1)$

 If $l-i=2^d+2u2^d, k-j=2^e+2v2^e$, without loss of
generality, assume $d<e$, by Lemma 2.2, $L(s) = 2^n-2^d,d>0$.

Note that $k-i=2c+1$, by Lemma 2.4, if $d=e$, then $L(s)=2^n-(2^d+1)$.

Based on 6 cases above, we conclude that the lemma is true.
\end{proof}\

With above important result, we have the following result with
constraint on the position of nonzero elements.

\noindent {\bf Corollary   2.1} Suppose that $s$ is a binary sequence
with period $2^n$ and there are only 4 non-zero elements, and $s$ can
be decomposed into the superposition of  $E_{ij}$ and  $E_{kl}$. If non-zero positions of
 $E_{ij}$ are $i$ and $j, j-i$ is an odd number, and non-zero
positions of  $E_{kl}$ are $k$ and $l, l-k$ is also an odd number, and $i<k, k-i=4c+2, |l-j|=4d+2$, or $|k-j|=4c+2, |l-i|=4d+2$,
then the linear complexity is $2^n-3$.

\begin{proof}\
 According to case 1), 3) and 5) of Lemma 2.5, if $k-i=4c+2, |l-j|=4d+2$, then $|l-j|=2+4d, k-i=2+4c$.
  By Lemma 2.4, noting that $j-i=2a+1$, so $L(s) = 2^n-(2+1)$.

According to case 2), 4) and 6) of Lemma 2.5, if $|k-j|=4c+2,
|l-i|=4d+2$, then it is easy to know that $k-i$ is odd, thus
$|k-j|=2+4c, |l-i|=2+4d$. By Lemma 2.4, $L(s) = 2^n-(2+1)$.
\end{proof}\

Alternatively, if $E_{ij}$ and  $E_{kl}$ have linear complexity of
$2^n-2$, we will have the following result.

\noindent {\bf Corollary   2.2} If  $s$ is a binary sequence with
period $2^n$ and there are only 4 non-zero elements, and $s$ can be
decomposed into the sum of two $E_{ij}$, in which each has linear
complexity $2^n-2$, then the linear complexity of $s$ is
$2^n-(2^d+1)2^e,e=0,1, d>0$ or $2^n-2^d, d>1$.

\begin{proof}\
 Suppose that non-zero positions of the first $E_{ij}$ are $i$ and $j$, $j-i=4a+2$,
 and non-zero positions of the second $E_{ij}$ are k and $l, l-k=4b+2$, where $i<k$.

If $k-i=2c+1$, according to Lemma 2.4, then $L(s) = 2^n-(2+1)$.

If $k-i=2c$, the corresponding polynomial of $E_i+E_j +E_k+E_l$ is
given by

$x^i+x^j+x^k+x^l=x^i(1+x^{j-i}+x^{k-i}+x^{l-k+k-i})$

Therefore, we only need to consider

$1+x^{j-i}+x^{k-i}+x^{l-k+k-i}=1+(x^2)^{2a+1}+(x^2)^c+(x^2)^{2b+1+c}=1+y^{2a+1}+y^c+y^{2b+1+c}$

According to Lemma 2.5, $L(s) = 2^n-(2^d+1)2, d>0$ or $2^n-2^d, d>1$.
\end{proof}\

Now we can obtain the following conclusions according to Lemma 2.5
and Corollary 2.2.

\noindent {\bf Proposition 2.2} Suppose that $s$ is a binary
sequence with period $2^n$ and there are four non-zero elements,
then the necessary and sufficient conditions for the linear
complexity of $s$ being $2^n-3$ are as follows. (i) $s$ can be
decomposed into the superposition of $E_{ik}$ and  $E_{jl}$, in
which each has linear complexity $2^n-2$; (ii) if non-zero positions
of $E_{ik}$ are $i$ and $k$, with $k-i=4c+2$, and non-zero positions
of the second $E_{jl}$ are $j$ and $l$, with $l-j=4d+2$, where
$i<j$, then $j-i=2a+1$(or $|l-k|=2b+1$ or $|l-i|=2e+1$ or
$|k-j|=2f+1$).

 \unitlength=0.04350mm
\begin{picture}(2200,1200)(-80,-50)

\drawcenteredtext{130}{1000}{$k$} \drawcenteredtext{1070}{1000}{$l$}
\drawcenteredtext{600}{1070}{$2b+1$}

\drawcenteredtext{130}{200}{$i$} \drawcenteredtext{1070}{200}{$j$}
\drawcenteredtext{600}{130}{$2a+1$}

\drawpath{200}{1000}{1000}{1000}
\drawpath{1000}{1000}{1000}{200}
\drawcenteredtext{1070}{600}{$4d+2$}
\drawpath{200}{1000}{1000}{200} \drawcenteredtext{400}{800}{$2f+1$}
\drawpath{200}{1000}{200}{200} \drawcenteredtext{100}{600}{$4c+2$}

\drawpath{200}{200}{1000}{200}
\drawpath{1000}{1000}{200}{200} \drawcenteredtext{800}{800}{$2e+1$}
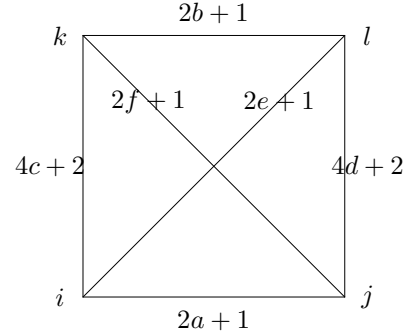
\drawcenteredtext{700.0}{0}{Figure 2.1 A graphic illustration of
Proposition 2.2}
\end{picture}

The above result gives a necessary and sufficient condition for a
sequence with linear complexity $2^n-3$ by using the proposed
decomposition. It seems that this relationship can be manipulated
recursively. Before we investigate it further, we can also
illustrate this with a graph in Figure 2.1. The only 4 non-zero
positions of sequence $s$
 are $i, j, k$ and $l$. As $k-i=4c+2$, $l-j=4d+2$, and $j-i=2a+1$, so $l-k=l-j+j-i-(k-i)=4d+2+2a+1-(4c+2)$ is odd.
 With such cube illustration, we can obtain a result on the
stable $2-error$ linear complexity for a periodic sequence.

\noindent {\bf Proposition 2.3} Suppose that $s$ is a binary sequence with
period $2^n$ and its Hamming weight is even, then the maximum stable
2-error linear complexity of $s$ is $2^n-3$.

\begin{proof}\ Assume that $L(s) = 2^n-1$, then $s$ can be decomposed into the sum of several $E_{ij}$ and
the number of $E_{ij}$ with linear complexity $2^n-1$ is odd.
According to Lemma 2.2, if an $E_{ij}$ with linear complexity
$2^n-1$ is removed, then the linear complexity of $s$ will be less
than $2^n-1$, namely the 2-error linear complexity of $s$ is less than
$2^n-1$.

Assume that $L(s) = 2^n-2$, then $s$ can be decomposed into the sum of
several $E_{ij}$ and the number of $E_{ij}$ with linear complexity
$2^n-2$ is odd. If an $E_{ij}$ with linear complexity $2^n-2$ is
removed, then the linear complexity of $s$ will be less than $2^n-2$,
namely the 2-error linear complexity of $s$ is less than $2^n-2$.

Assume that $L(s) = 2^n-3$, without loss of generality, here we only
discuss the case that $s$ has 4 non-zero elements: $e_i, e_j, e_k$ and
$e_l$, and $L(E_i+E_j +E_k+E_l)= 2^n-3$. If any two of them are
removed, by Proposition 2.2, the linear complexity of remaining elements
of the sequence is $2^n-1$ or $2^n-2$. From Figure 2.1, after $e_i$
and $e_l$ are changed to zero, we can see that the linear complexity of the
sequence composed by $e_j$ and $e_k$ is $2^n-1$.

If the position of one element from $e_i, e_j, e_k$ and $e_l$ is
changed, then there exist two elements, of which the position
difference remains unchanged as odd, thus $L(s) \ge 2^n-3$ .

If two nonzero elements are added to the position outside $e_i, e_j,
e_k$ and $e_l$, namely an $E_{ij}$ with linear complexity $2^n-2^d$
is added to sequence $s$, according to Lemma 2.2, the linear
complexity will be $2^n-1$, $2^n-2$ or $2^n-3$.

The proof is completed.
\end{proof}\

Next, we present an example to illustrate Proposition 2.3.

\noindent {\bf Example 2.1} Given the following three sequences,
11110$\cdots$0 with linear complexity $2^n-3$; 01010$\cdots$0 or
10100$\cdots$0 with linear complexity $2^n-2$ and 01100$\cdots$0 or
10010$\cdots$0 with linear complexity $2^n-1$. If two additional
nonzero elements are added to 11110$\cdots$0, namely an $E_{ij}$
whose linear complexity is $2^n-2^d$ is added to it, according to
Lemma 2.2, the linear complexity of the produced sequence will
become $2^n-1$, $2^n-2$ or $2^n-3$.

For instance, suppose that 1110$\cdots$010$\cdots$0 is the
addition of 11110$\cdots$0 and 0001$\cdots$010$\cdots$0.
 We here only consider the case that the position difference of the last
two nonzero elements is $2c+1$. According to case 5) of Lemma 2.5,
$j-i=1,l-k=2c+1$, so $k-j=1, l-j=2(c+1)$.

Noted that $k-i=2$, if $l-j=2^d(2u+1)$,  according to Lemma 2.2, $L(s) = 2^n-2$
when $d>1$.

 If $d=1$, since $j-i=1$, according to Lemma 2.4, $L(s) = 2^n-3$.

In all cases, the linear complexity is less than $2^n-3$.

\section{Cube Theory and Main Results}

Before presenting main results, we first give a special case.

\noindent {\bf Lemma  3.1} Suppose that $s$ is a binary sequence with
period $2^n$ and there are 8 non-zero elements, thus $s$ can be
decomposed into the superposition of  $E_{ij}$, $E_{kl}$,  $E_{mn}$ and  $E_{pq}$.
 Suppose that non-zero
positions of  $E_{ij}$ are $i$ and $j, j-i=2a+1$, and
non-zero positions of  $E_{kl}$ are $k$ and $l, l-k=2b+1$,
and $k-i=4c+2, l-j=4d+2$, and non-zero positions of
$E_{mn}$ are $m$ and $n$, non-zero positions of $E_{pq}$ are
$p$ and $q$, and $m-i=4+8u, n-j=4+8v, p-k=4+8w, q-l=4+8y$, where
$a,b,c,d,u,v,w$ and $y$ are all non-negative integers, then the
linear complexity of $s$ is $2^n-7$.

\begin{proof}\
According to Corollary 2.1, $L(E_i+E_j+E_k+E_l)=2^n-3$.

As $m-n=m-i-(n-j)-(j-i)$, $p-q=p-k-(q-l)-(l-k)$, thus both $m-n$ and $p-q$ are odd numbers.

As $p-m=p-k-(m-i)+(k-i)$, $q-n=q-l-(n-j)+(l-j)$, thus both $p-m$ and $q-n$ are
  multiples of 2, but not  multiples of 4. According to Corollary 2.1, $L(E_m+E_n+E_p+E_q)=2^n-3$.

Similar to the proof of Lemma 2.4 \cite{Zhou_Liu}, the corresponding polynomial of
$E_i+E_k+E_m+E_p$ is given by
\begin{eqnarray*}
&&x^i+x^k+x^m+x^p\\
&=&x^i(1-x^{4})[(1+x^{4}+x^{2\cdot4}+\cdots +x^{2u\cdot4})\\
&&\ \ \ \ \ +x^{k-i}(1+x^{4}+x^{2\cdot4}+\cdots +x^{2w\cdot4})]\\
&=&x^i(1-x^{4})[1+x^{k-i}+(x^{4}+x^{2\cdot4}+\cdots +x^{2u\cdot4})\\
&&\ \ \ \ \ +x^{k-i}(x^{4}+x^{2\cdot4}+\cdots +x^{2w\cdot4})]\\
&=&x^i(1-x^{4})[1+x^{4c+2}+(x^{4}+x^{2\cdot4}+\cdots +x^{2u\cdot4})\\
&&\ \ \ \ \ +x^{k-i}(x^{4}+x^{2\cdot4}+\cdots +x^{2w\cdot4})]\\
&=&x^i(1-x)^{6}[(1+x^2+x^{4}+\cdots+x^{4c})\\
&&\ \ \  +(x^{4}+x^{3\cdot4}+\cdots +x^{(2u-1)\cdot4})(1+x)^{2}\\
&&\ \ \  +x^{k-i}(x^{4}+x^{3\cdot4}+\cdots
+x^{(2w-1)\cdot4})(1+x)^{2}]
 \end{eqnarray*}

The corresponding polynomial of $E_j+E_l+E_n+E_q$ is given by
\begin{eqnarray*}
&&x^j+x^l+x^n+x^q\\
&=&x^j(1-x^{4})[(1+x^{4}+x^{2\cdot4}+\cdots +x^{2v\cdot4})\\
&&\ \ \ \ \ +x^{l-j}(1+x^{4}+x^{2\cdot4}+\cdots +x^{2y\cdot4})]\\
&=&x^j(1-x)^{6}[(1+x^2+x^{4}+\cdots+x^{4d})\\
&&\ \ \  +(x^{4}+x^{3\cdot4}+\cdots +x^{(2v-1)\cdot4})(1+x)^{2}\\
&&\ \ \  +x^{l-j}(x^{4}+x^{3\cdot4}+\cdots
+x^{(2y-1)\cdot4})(1+x)^{2}]
 \end{eqnarray*}

The corresponding polynomial of $E_i+E_j+E_k+E_l+E_m+E_n+E_p+E_q$ is
given by
\begin{eqnarray*}
&&x^i+x^j+x^k+x^l+x^m+x^n+x^p+x^q\\
&=&x^i(1-x)^{6}\{(1+x^2+x^{4}+\cdots+x^{4c})\\
&&\ \ \  +(x^{4}+x^{3\cdot4}+\cdots +x^{(2u-1)\cdot4})(1+x)^{2}\\
&&\ \ \  +x^{k-i}(x^{4}+x^{3\cdot4}+\cdots
+x^{(2w-1)\cdot4})(1+x)^{2}\\
&&\ \ \ +x^{j-i}[(1+x^2+x^{4}+\cdots+x^{4d})\\
&&\ \ \  +(x^{4}+x^{3\cdot4}+\cdots +x^{(2v-1)\cdot4})(1+x)^{2}\\
&&\ \ \  +x^{l-j}(x^{4}+x^{3\cdot4}+\cdots
+x^{(2y-1)\cdot4})(1+x)^{2}]\}\\
&=&x^i(1-x)^{6}\{1+x^{j-i}+(x^2+x^{4}+\cdots+x^{4c})\\
&&\ \ \  +(x^{4}+x^{3\cdot4}+\cdots +x^{(2u-1)\cdot4})(1+x)^{2}\\
&&\ \ \  +x^{k-i}(x^{4}+x^{3\cdot4}+\cdots
+x^{(2w-1)\cdot4})(1+x)^{2}\\
&&\ \ \ +x^{j-i}[(x^2+x^{4}+\cdots+x^{4d})\\
&&\ \ \  +(x^{4}+x^{3\cdot4}+\cdots +x^{(2v-1)\cdot4})(1+x)^{2}\\
&&\ \ \  +x^{l-j}(x^{4}+x^{3\cdot4}+\cdots
+x^{(2y-1)\cdot4})(1+x)^{2}]\}\\
&=&x^i(1-x)^{7}\{1+x+x^{2}+\cdots+x^{2a}\\
&&\ \ \ +x^2(1+x)(1+x^{4}+\cdots+x^{4(c-1)})\\
&&\ \ \  +(x^{4}+x^{3\cdot4}+\cdots +x^{(2u-1)\cdot4})(1+x)\\
&&\ \ \  +x^{k-i}(x^{4}+x^{3\cdot4}+\cdots
+x^{(2w-1)\cdot4})(1+x)\\
&&\ \ \ +x^{j-i}[x^2(1+x)(1+x^{4}+\cdots+x^{4(d-1)})\\
&&\ \ \  +(x^{4}+x^{3\cdot4}+\cdots +x^{(2v-1)\cdot4})(1+x)\\
&&\ \ \  +x^{l-j}(x^{4}+x^{3\cdot4}+\cdots
+x^{(2y-1)\cdot4})(1+x)]\}
 \end{eqnarray*}

 The number of items in $(1+x+x^{2}+\cdots+x^{2a})$  is odd, thus
 there is no factor $(1 + x)$ in $(1+x+x^{2}+\cdots+x^{2a})$.

 $$\gcd((1-x)^{2^n},x^i+x^j+x^k+x^l+x^m+x^n+x^p+x^q)=(1-x)^7$$

 It is followed by $L(s)=2^n-7$.
\end{proof}\

\unitlength=0.26mm

\begin{picture}(320,320)(18,-5)

\drawcenteredtext{70}{280}{$p$} \drawcenteredtext{45}{250}{$2$}
\drawcenteredtext{160}{280}{$1$}

\drawcenteredtext{250}{280}{$q$} \drawcenteredtext{250}{200}{$4$}
\drawcenteredtext{225}{250}{$2$}

 \drawpath{80}{280}{240}{280}

\drawpath{80}{280}{30}{220}
\drawcenteredtext{20}{220}{$m$} \drawcenteredtext{20}{140}{$4$}

\drawcenteredtext{110}{220}{$1$}

\drawcenteredtext{200}{220}{$n$} \drawcenteredtext{200}{140}{$4$}

\drawpath{30}{220}{190}{220}

\drawpath{190}{220}{240}{280}

\drawpath{30}{220}{240}{280}

\drawcenteredtext{20}{60}{$i$} \drawcenteredtext{200}{60}{$j$}

\drawcenteredtext{110}{60}{$1$}

 \drawpath{30}{60}{190}{60}
\drawpath{30}{60}{30}{220}
\drawpath{190}{60}{190}{220}

\drawcenteredtext{250}{120}{$l$}\drawcenteredtext{225}{170}{$2$}
\drawcenteredtext{225}{90}{$2$}

\drawpath{190}{60}{240}{120}

\drawpath{190}{220}{240}{120}

\drawpath{190}{220}{80}{280}

\drawpath{240}{280}{240}{120}

\drawcenteredtext{70}{120}{$k$} \drawcenteredtext{45}{90}{$2$}
\drawcenteredtext{160}{120}{$1$}


 \drawpath{30.00}{60.00}{35.00}{66.00}
 \drawpath{36.25}{67.50}{40.00}{72.00}
 \drawpath{41.25}{73.50}{45.00}{78.00}
 \drawpath{46.25}{79.50}{50.00}{84.00}
 \drawpath{51.25}{85.50}{55.00}{90.00}
 \drawpath{56.25}{91.50}{60.00}{96.00}
 \drawpath{61.25}{97.50}{65.00}{102.00}
 \drawpath{66.25}{103.50}{70.00}{108.00}
 \drawpath{71.25}{109.50}{75.00}{114.00}
 \drawpath{76.25}{115.50}{80.00}{120.00}


 \drawpath{80.00}{120.00}{96.00}{120.00}
 \drawpath{100.00}{120.00}{112.00}{120.00}
 \drawpath{116.00}{120.00}{128.00}{120.00}
 \drawpath{132.00}{120.00}{144.00}{120.00}
 \drawpath{148.00}{120.00}{160.00}{120.00}
 \drawpath{164.00}{120.00}{176.00}{120.00}
 \drawpath{180.00}{120.00}{192.00}{120.00}
 \drawpath{196.00}{120.00}{208.00}{120.00}
 \drawpath{212.00}{120.00}{224.00}{120.00}
 \drawpath{228.00}{120.00}{240.00}{120.00}


 \drawpath{80.00}{120.00}{80.00}{136.00}
 \drawpath{80.00}{140.00}{80.00}{152.00}
 \drawpath{80.00}{156.00}{80.00}{168.00}
 \drawpath{80.00}{172.00}{80.00}{184.00}
 \drawpath{80.00}{188.00}{80.00}{200.00}
 \drawpath{80.00}{204.00}{80.00}{216.00}
 \drawpath{80.00}{220.00}{80.00}{232.00}
 \drawpath{80.00}{236.00}{80.00}{248.00}
 \drawpath{80.00}{252.00}{80.00}{264.00}
 \drawpath{80.00}{268.00}{80.00}{280.00}


 \drawpath{80.00}{120.00}{91.00}{114.00}
 \drawpath{93.75}{112.50}{102.00}{108.00}
 \drawpath{104.75}{106.50}{113.00}{102.00}
 \drawpath{115.75}{100.50}{124.00}{96.00}
 \drawpath{126.75}{94.50}{135.00}{90.00}
 \drawpath{137.75}{88.50}{146.00}{84.00}
 \drawpath{148.75}{82.50}{157.00}{78.00}
 \drawpath{159.75}{76.50}{168.00}{72.00}
 \drawpath{170.75}{70.50}{179.00}{66.00}
 \drawpath{181.75}{64.50}{190.00}{60.00}


 \drawpath{80.00}{120.00}{75.00}{130.00}
 \drawpath{73.75}{132.50}{70.00}{140.00}
 \drawpath{68.75}{142.50}{65.00}{150.00}
 \drawpath{63.75}{152.50}{60.00}{160.00}
 \drawpath{58.75}{162.50}{55.00}{170.00}
 \drawpath{53.75}{172.50}{50.00}{180.00}
 \drawpath{48.75}{182.50}{45.00}{190.00}
 \drawpath{43.75}{192.50}{40.00}{200.00}
 \drawpath{38.75}{202.50}{35.00}{210.00}
 \drawpath{33.75}{212.50}{30.00}{220.00}



 \drawpath{30.00}{60.00}{51.00}{66.00}
 \drawpath{56.25}{67.50}{72.00}{72.00}
 \drawpath{77.25}{73.50}{93.00}{78.00}
 \drawpath{98.25}{79.50}{114.00}{84.00}
 \drawpath{119.25}{85.50}{135.00}{90.00}
 \drawpath{140.25}{91.50}{156.00}{96.00}
 \drawpath{161.25}{97.50}{177.00}{102.00}
 \drawpath{182.25}{103.50}{198.00}{108.00}
 \drawpath{203.25}{109.50}{219.00}{114.00}
 \drawpath{224.25}{115.50}{240.00}{120.00}

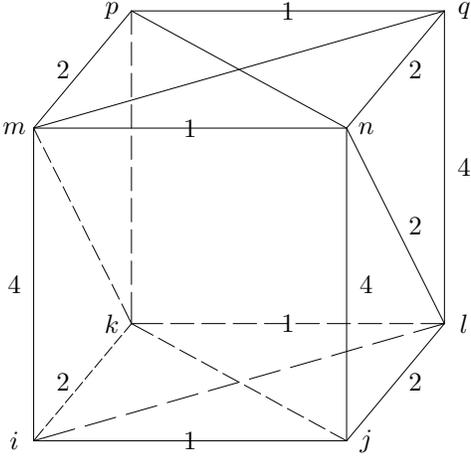
\drawcenteredtext{150.0}{10}{Figure 3.1 A graphic illustration of
Lemma 3.1 }
\end{picture}

For the convenience of presentation, we introduce some definitions.

\noindent {\bf Definition  3.1} Suppose that the difference of
positions of two non-zero elements of sequence $s$ is $(2x+1)2^y$,
both $x$ and $y$ are non-negative integers, then the distance between
the two elements is defined as  $2^y$.

\noindent {\bf Definition  3.2} Suppose that $s$ is a binary sequence
with period $2^n$, and there are $2^m$ non-zero elements in $s$, and
$0\le i_1< i_2<\cdots<i_m<n$. If $m=1$, then there are 2 non-zero
elements in $s$ and the distance between the two elements is
$2^{i_1}$, so it is called as a 1-cube. If $m = 2$, then $s$ has 4
non-zero elements which form a rectangle, the lengths of 4 sides are
$2^{i_1}$ and $2^{i_2}$ respectively, so it is called as a 2-cube.
In general, $s$ has $2^{m-1}$ pairs of non-zero elements, in which
there are $2^{m-1}$ non-zero elements which form a $(m-1)$-cube, the
other $2^{m-1}$ non-zero elements also form a $(m-1)$-cube, and the
distance between each pair of elements are all  $2^{i_m}$, then the
sequence $s$ is called as an $m$-cube, and the linear complexity of $s$ is
 called as the linear complexity of the cube as well.

\noindent {\bf Definition  3.3}
 A non-zero element  of sequence $s$ is  called a vertex.
Two vertexes can  form an edge. If the distance between the two
elements (vertices) is   $2^y$, then the length of the edge is
defined as $2^y$.

Now we consider the linear complexity of a sequence with only one cube.

\noindent {\bf Theorem   3.1} Suppose that $s$ is a binary sequence
with period $2^n$, and non-zero elements of $s$ form an $m$-cube, if
lengths of  edges are $ i_1, i_2,\cdots ,i_m$ $(0\le i_1<
i_2<\cdots<i_m<n )$ respectively, then
$L(s)=2^n-(2^{i_1}+2^{i_2}+\cdots+2^{i_m})$.

\begin{proof}\

Similar to the proof of Lemma 3.1, it is easy to prove Theorem   3.1
with mathematical induction.

Based on Algorithm 2.1, we give another proof
from a different perspective.
In the $k$th step, $1\le k\le n$, if and only if one period of the sequence can not be divided into two equal parts, then the
 linear complexity should be increased by half period. In the $k$th step, the
 linear complexity can be increased by maximum $2^{n-k}$.

Suppose that non-zero elements of sequence $s$ form a $m$-cube,
lengths of  edges are $ i_1, i_2,\cdots ,i_m$ $(0\le i_1<
i_2<\cdots<i_m<n )$ respectively. Then in the $(n-i_m)$th step, one
period of the sequence can  be divided into two equal parts, then
the
 linear complexity should not be increased by $2^{i_m}$.

$\cdots\cdots$

 In the $(n-i_2)$th step, one period of the sequence can  be divided into two equal parts, then the
 linear complexity should not be increased by $2^{i_2}$.

 In the $(n-i_1)$th step, one period of the sequence can  be divided into two equal parts, then the
 linear complexity should not be increased by $2^{i_1}$.

Therefore, $L(s)=1+1+2+2^2+\cdots+2^{n-1}-(2^{i_1}+2^{i_2}+\cdots+2^{i_m})=2^n-(2^{i_1}+2^{i_2}+\cdots+2^{i_m})$.

The proof is completed now.
\end{proof}\

There is a 3-cube in Figure 3.1. $L(s)=2^n-(1+2+4)$, and lengths of
edges are $1,2$ and 4 respectively. Next we give a decomposition
result.

\noindent {\bf Theorem   3.2} Suppose that $s$ is a binary sequence
with period $2^n$, and $L(s)=2^n-(2^{i_1}+2^{i_2}+\cdots+2^{i_m})$,
where $0\le i_1< i_2<\cdots<i_m<n$, then the sequence $s$ can be
decomposed into several disjoint cubes, and only one cube has the
linear complexity $2^n-(2^{i_1}+2^{i_2}+\cdots+2^{i_m})$, other
cubes possess distinct linear complexity which are all less than
$2^n-(2^{i_1}+2^{i_2}+\cdots+2^{i_m})$.

\begin{proof}\
The mathematical induction will be applied to the degree $d$ of
$s^N(x)$. For $d <3$, by Lemma 2.3, the theorem is obvious.

We first consider a simple case.

A) Suppose that
$L(s)=2^n-(2^{i_1}+2^{i_2}+\cdots+2^{i_m}+2^{i_{m+1}})$, and the
Hamming weight of $s$ is the minimum, namely
$L(s)\ne2^n-(2^{i_1}+2^{i_2}+\cdots+2^{i_m}+2^{i_{m+1}})$ when
remove 2 or more non-zero elements. Next we prove that $s$ consists
of one $(m+1)$-cube exactly. Let

\begin{eqnarray*}
s^N(x)&=&(1-x^{2^{i_1}})(1-x^{2^{i_2}})\cdots(1-x^{2^{i_m}})(1-x^{2^{i_{m+1}}})\\
&&\ \ \ [1+f(x)(1-x)]
\end{eqnarray*}

Then
$t^N(x)=(1-x^{2^{i_1}})(1-x^{2^{i_2}})\cdots(1-x^{2^{i_m}})[1+f(x)(1-x)]$
corresponds to a sequence $t$ whose linear complexity is
$L(t)=2^n-(2^{i_1}+2^{i_2}+\cdots+2^{i_m})$. The degree of $t^N(x)$
is less than the degree of $s^N(x)$, so the mathematical induction
can be applied.

In the following, we consider two cases.

1) The Hamming weight of sequence $t$ is $2^m$. By mathematical induction, $t$ is
an $m$-cube. Since  $s^N(x)=t^N(x)(1-x^{2^{i_{m+1}}})=t^N(x)+x^{2^{i_{m+1}}}t^N(x)$, and  $0\le
i_1< i_2<\cdots<i_m<i_{m+1}<n$, so $s$ is a $(m+1)$-cube and its Hamming
weight is $2^{m+1}$.

2)  The Hamming weight of sequence $t$ is $2^m+2y$. By mathematical induction,
the sequence $t$ can be
decomposed into several disjoint cubes, and only one cube has the
linear complexity $2^n-(2^{i_1}+2^{i_2}+\cdots+2^{i_m})$. Thus

$t^N(x)=(1-x^{2^{i_1}})(1-x^{2^{i_2}})\cdots(1-x^{2^{i_m}})[1+g(x)(1-x)+h(x)(1-x)]$,
and
$u^N(x)=(1-x^{2^{i_1}})(1-x^{2^{i_2}})\cdots(1-x^{2^{i_m}})[1+g(x)(1-x)]$,
corresponds to an $m$-cube, its non-zero elements form a set denoted
by A.

$v^N(x)=(1-x^{2^{i_1}})(1-x^{2^{i_2}})\cdots(1-x^{2^{i_m}})h(x)(1-x)$
corresponds to several cubes, whose 2y non-zero elements form a set
denoted by B.

Assume that  $b\in B, bx^{2^{i_{m+1}}}\in A$, we swap $b$ and
$bx^{2^{i_{m+1}}}$, namely let  $b\in A, bx^{2^{i_{m+1}}}\in B$. It
is easy to show that the linear complexity of the sequence to which
$u^N(x)$ corresponds remains unchanged. The new $u^N(x)$ is still an $m$-cube.

$s^N(x)=t^N(x)(1-x^{2^{i_{m+1}}})=u^N(x)+v^N(x)-u^N(x)x^{2^{i_{m+1}}}-v^N(x)x^{2^{i_{m+1}}}$,
$u^N(x)x^{2^{i_{m+1}}}$ corresponds to $2^m$ non-zero elements which
form a set denoted by C. $v^N(x)x^{2^{i_{m+1}}}$ corresponds to 2y
non-zero elements which form a set denoted by D.

By definition, set A and set C disjoint, set B and set D disjoint.

Suppose that  set A and set D intersects. Thus there exists  $b\in B$, such that $bx^{2^{i_{m+1}}}\in A$, which contradicts
 the assumption that  $b\in A, bx^{2^{i_{m+1}}}\in B$. So set A and set D disjoint.

As set A and set B disjoint, we know that set C and set D disjoint.

We now prove by contradiction that Set C and B disjoint. 

Suppose that $b\in B, b=ax^{2^{i_{m+1}}}\in C, a\in A$, then
$ax^{2(2^{i_{m+1}})}$ must be in D,
so sequence $s$ has non-zero elements $a$ and
$ax^{2(2^{i_{m+1}})}$. The linear complexity of the sequence with only  non-zero elements $a$ and
$ax^{2(2^{i_{m+1}})}$ is
$$2^n-2\cdot2^{i_{m+1}}<2^n-(2^{i_1}+2^{i_2}+\cdots+2^{i_m}+2^{i_{m+1}}).$$

By Lemma 2.2, if the two non-zero elements are changed to zero, the
linear complexity of $s$ remains unchanged. It contradicts the
assumption that the Hamming weight is the minimum, so $A$ and $C$
form a $(m+1)$-cube exactly, and its linear complexity is
$2^n-(2^{i_1}+2^{i_2}+\cdots+2^{i_m}+2^{i_{m+1}})$.

By the assumption of Case A), $s$ has minimum Hamming weight, so $s$
consists of a $(m+1)$-cube exactly.

B) Let $s^N(x)= u^N(x)+ v^N(x)$, where the Hamming weight of
$u^N(x)$ is the minimum, and
$$L(u)=2^n-(2^{i_1}+2^{i_2}+\cdots+2^{i_m}+2^{i_{m+1}}).$$ From Case A),
$u^N(x)$ consists of a $(m+1)$-cube exactly.

Let $v^N(x)= y^N(x)+ z^N(x)$, where the Hamming weight of $y^N(x)$
is minimum, and $L(y)=L(v)$. By Case A), $y^N(x)$ consists of only one cube
exactly. By analogy, we can prove that $s$ consists of several cubes,
and only one cube has the linear complexity of
$2^n-(2^{i_1}+2^{i_2}+\cdots+2^{i_m}+2^{i_{m+1}})$, other cubes
possess distinct linear complexity which are all less than
$2^n-(2^{i_1}+2^{i_2}+\cdots+2^{i_m}+2^{i_{m+1}})$.

This completes proof.
\end{proof}\

The following examples can help us understand the proof of Theorem
3.2.

\noindent {\bf Example 3.1} One can see that
$(1+x)(1+x^2)[1+x^5(1+x^2)]=1+x+x^2+x^3+x^5+x^6+x^9+x^{10}$
 and this polynomial corresponds to a sequence in which there are 8 non-zero elements.
 This sequence can be decomposed into two 2-cube: $(1+x)(1+x^2)$ and
 $(1+x)(1+x^4)x^5$. On the other hand, $(1+x)(1+x^2)[1+x^5(1+x^2)](1+x^4)=1+x+x^2+x^3+x^4+x^7+x^{13}+x^{14}$
and this higher degree polynomial corresponds to a sequence in which
there are also 8 non-zero elements, which can be decomposed into
only one 3-cube with linear complexity of $2^n-(1+2+4)$, and the
lengths of edges 1, 2 and 4 respectively. This indicates that after
polynomial product, the non zero elements are not increased.

Suppose that the linear complexity of $s$ can reduce when at least
$k$ elements of $s$ are changed. By Lemma 2.2, the linear complexity
of the binary sequence, in which elements at exactly those $k$
positions are all nonzero, must be $L(s)$. According to Theorem 3.1
and Theorem 3.2, it is easy to get the following conclusion.

\noindent {\bf Corollary   3.1} Suppose that $s$ is a binary sequence
with period  $2^n$, and $L(s)=2^n-(2^{i_1}+2^{i_2}+\cdots+2^{i_m})$,
where  $0\le i_1< i_2<\cdots<i_m<n$. If $k_{\min}$ is the minimum,
such that $k_{\min}$-error linear complexity is less than $L(s)$, then
$k_{\min}=2^m$.

Corollary 3.1 was first proved by Kurosawa et al. \cite{Kurosawa},
and later it was proved by Etzion et al. \cite{Etzion} with
different approaches. Here we obtain this result from the cube
theory and different from the previous approaches.

Consider a $k$-cube, if lengths of edges are 1,2,$2^2,\cdots,$ and
$2^{k-1}$ respectively, and the linear complexity is $2^n-(2^k-1)$.
By Theorem 3.1 and Theorem 3.2, we can obtain the following results
on stability.

\noindent {\bf Corollary   3.2} Suppose that $s$ is a binary sequence
with period  $2^n$ and its Hamming weight is even, then the maximum
stable $2^{k-1},\cdots, (2^k-2)$ or $(2^k-1)$-error linear complexity of
s are all $2^n-(2^k-1)(k>0)$.

The following is an example to illustrate Corollary 3.2.

Let $s$ be the binary sequence $\overbrace{11\cdots11}^{2^k}0\cdots0$. Its period is
$2^n$, and there are only $2^k$ continuous nonzero elements at the
beginning of the sequence. Then it is a $k$-cube, and the
$2^{k-1},\cdots, (2^k-2)$ or $(2^k-1)$-error linear complexity of $s$ are
all $2^n-(2^k-1)$.

After at most $e(0\le e\le 2^k-1)$ elements of a period in the above
sequence are changed, the linear complexity of all new sequences are
not decreased, so the
original sequence possesses stable e-error linear complexity.

According to Lemma 2.2, if a sequence whose linear complexity is
less than $2^n-(2^k-1)$ is added to the sequence with linear
complexity $2^n-(2^k-1)$, then the linear complexity of the new
sequence is still $2^n-(2^k-1)$, and the $2^{k-1},\cdots, (2^k-2)$ or
$(2^k-1)$-error linear complexity of the new sequence are all
$2^n-(2^k-1)$.

By combining Corollary   3.1 and Corollary   3.2,  we can achieve
the following theorem.

\noindent {\bf Theorem   3.3} For $ 2^{l-1}\le k<2^{l}$, there exists a $2^n$-periodic  binary sequence  $s$
with stable $k$-linear complexity $2^n-(2^l-1)$, such that
$$L_k(s)=\max\limits_tL_k(t)$$ where   $t$ is any
$2^n$-periodic  binary sequence.

It is reminded that CELCS (critical error linear complexity
spectrum) is studied by Etzion et al. \cite{Etzion}. The CELCS of the
sequence $s$ consists of the ordered set of points $(k,c_k(s))$
satisfying $c_k(s)> c_{k'}(s)$, for $k'>k$; these are the points
where a decrease occurs in the $k$-error linear complexity, and thus
are called critical points.

Let  $s$  be a binary sequence whose period is $2^n$ and it has only
one $m$-cube. Then $s$ has only two critical points: $(0,l(s)),
(2^m,0)$.

In the following we study binary sequences which consist of several
cubes and the $t$th decrease in the $k$-error linear complexity, where $t>1$.
By Theorem 3.2, if $s$ is a $2^n$-periodic binary sequence,
then it can be decomposed into several disjoint cubes. The following
examples show that the cube decomposition of a sequence is not
unique.

\noindent {\bf Example 3.2} $1+x+x^3+x^4+x^7+x^8$ can be decomposed into a 1-cube $1+x$, whose linear complexity is $2^n -1$,
  and a 2-cube  $x^3+x^4+x^7+x^8$, whose linear complexity is $2^n -(1+4)$.

It can also be decomposed into a 1-cube $x^3+x^4$, whose linear
complexity is $2^n -1$, a 1-cube $x+x^7$, whose linear complexity is
$2^n -2$, and another 1-cube $1+x^8$, whose linear complexity is
$2^n -8$.

It can also be decomposed into a 1-cube $x^7+x^8$, whose linear
complexity is $2^n -1$, a 1-cube $x+x^3$, whose linear complexity is
$2^n -2$, and another 1-cube $1+x^4$, whose linear complexity is
$2^n -4$.

It can also be decomposed into a 1-cube $1+x^3$, whose linear
complexity is $2^n -1$, a 1-cube $x+x^7$, whose linear complexity
$2^n -2$, and another 1-cube $x^4+x^8$, whose linear complexity is
$2^n -4$.

$\cdots \cdots$

In fact, we do not know how many possible ways for such
decomposition. However, based on  Algorithm 2.1,
we may have a standard cube decomposition of sequence.

\noindent {\bf Algorithm 3.1}

\noindent {\bf Input:} $s^{(n)}$ is a binary sequence with period
$2^n$.

\noindent {\bf Output:} A  cube decomposition of sequence $s^{(n)}$.

\noindent Step 1. Let $s^{(n)}=[Left(s^{(n)}),Right(s^{(n)})]$.

\noindent Step 2. If $Left(s^{(n)})=Right(s^{(n)})$, then we consider $Left(s^{(n)})$. $Left(s^{(n)})$
is still a set of cubes, but the dimension of every cube reduced by 1.

\noindent Step 3. If $Left(s^{(n)})\neq Right(s^{(n)})$, then we
consider $Left(s^{(n)})\bigoplus Right(s^{(n)})$. Some cubes of  $s$
may be removed.  With these cubes removed recursively, we will
obtain a series of cubes in the ascending order of linear
complexity; while with these cubes recursively left, we will obtain
a series of cubes in the ascending order of linear complexity.

\noindent Step 4. Finally, by restoring the dimension reduced of
cubes, one can obtain a  series of cubes in the ascending order of
linear complexity.

 Obviously, this is a
 cube decomposition of sequence $s$. We define it as {\bf the standard cube decomposition} of sequence $s$.

Based on Algorithm 2.1 and the standard cube decomposition, it is
easy to prove the following result.

\noindent {\bf Proposition 3.1}
 Suppose that a complete graph  consists of all non-zero elements of
 sequence $s$, and the length of the edge $e$ is the maximum.
 Then edge $e$ must be in the smallest cube in terms of  linear complexity. Similarly,
 the smallest cube of the standard cube decomposition has the minimum linear complexity
 compared with  the smallest cube of other possible decompositions.

Next we use  Example 3.2 to illustrate the decomposition process. As
$1+x+x^3+x^4+x^7+x^8$ can be considered as a sequence 1101 1001 1000
0000

As $Left\neq Right$, then we consider $Left\bigoplus Right$. Then
the cube $1+x^8$ is removed.

Recursively, as $Left\neq Right$, then we consider $Left\bigoplus
Right$. This time the cube $x^3+x^7$ is removed. Only cube $x+x^4$
is retained. So the standard cube decomposition of
$1+x+x^3+x^4+x^7+x^8$ is $\{x+x^4$,  $x^3+x^7, 1+x^8\}$.

Obviously, the  cube $1+x^8$ has the minimum linear complexity among
all other decompositions.

Now we consider dependence relationship among different cubes.

In order to achieve the maximal decrease of the linear complexity of
a new sequence by superposing another sequence over the original
one, according to Lemma 2.2, a direct method is, if possible, to the
linear complexity of the first cube and let it be the same as the
linear complexity of the second cube. For Example 3.2, for the
polynomial $1+x+x^3+x^4+x^7+x^8$ with standard decomposition
$\{x+x^4$,  $x^3+x^7, 1+x^8\}$,  in order to make the linear
complexity of $x+x^4$ to be the same as $x^3+x^7$, we add
$x^{4}+x^{5}$ and obtain $x+x^{5}$, which has the same linear
complexity of $x^3+x^7$. Therefore,  we have that the critical
points of $1+x+x^3+x^4+x^7+x^8$ is
$(0,2^n-1)$,$(2,2^n-(2+4))$,$(4,2^n-8)$,$(6,0)$.

 To further investigate the critical point issue, we consider another example.

  \noindent {\bf Example 3.3} Consider $1+x^3+x^4+x^6+x^9+x^{11}+x^{12}+x^{14}$. Its standard cube decomposition is
 $\{1+x^9,x^3+x^{11},x^4+x^6+x^{12}+x^{14}\}$.
 If we change  $1+x^9$ to $x+x^9$, then  $x+x^3+x^4+x^6+x^9+x^{11}+x^{12}+x^{14}$ is a 3-cube with the linear complexity  $2^n-(1+2+8)$. So, both 2-error linear complexity and 4-error linear complexity of $1+x^3+x^4+x^6+x^9+x^{11}+x^{12}+x^{14}$ are all $2^n-(1+2+8)$. 6-error linear complexity is $2^n-(2+4+8)$.

 From above examples, we can find that  one cube change  may affect other cubes in the cube decomposition. This phenomena make the critical points detection more difficult in general.
 In order to discuss the critical points easily, we give the following concept.  If the change of one cube has an impact on other cubes, then the cube decomposition is
 defined as {\bf power relation}. In Example 3.2, the change of $x+x^4$ to  $x+x^5$ only has impact on one cube $x^3+x^7$, so  the cube decomposition is
 defined as {\bf first order power relation}.
In Example 3.3, the change of $1+x^9$ to  $x+x^9$  has impact on two cubes, so  the cube decomposition is
 defined as {\bf second order power relation}.

It is easy to verify the following, which provides a general
solution to find critical points of a $2^n$-periodic binary
sequence.

\noindent {\bf Theorem  3.4}  Suppose that $s$ is a binary sequence
with period  $2^n$, and $s$ has a unique cube decomposition without the $t$th order power relation, where $t>1$.
If the cubes are in descending order of   linear complexity, and their dimensions are $m_1, m_2,\cdots,m_t$, respectively, then $k$-error linear complexity will decrease when $k$ is $2^{m_1}$,$2^{m_1}+2^{m_2},\cdots, 2^{m_1}+2^{m_2}+\cdots+2^{m_t}$.

It is fair to say that the  requirement that sequence $s$ has a unique cube decomposition in
Theorem 3.4 is mild. Now we give a sufficient condition for sequence $s$ to have a unique cube decomposition.

\noindent {\bf Proposition 3.2}
Suppose that sequence $s$ can be decomposed into several disjoint cubes with distinct linear complexity.
If the minimum length of edges in all cubes of a binary sequence $s$ is $2^w$, and the distance between any two cubes is less than $2^w$, then sequence $s$  has a unique cube decomposition.

For example, $1+x^2+x^4+x^6+x^7+x^{15}$ can be decomposed into a 2-cube $1+x^2+x^4+x^6$, whose linear complexity is $2^n -(2+4)$,
  and a 1-cube  $x^7+x^{15}$, whose linear complexity is $2^n -8$. As the minimum length of edges in all cubes is $2$ and the distance between  two cubes is 1 less than 2, so $1+x^2+x^4+x^6+x^7+x^{15}$  has a unique cube decomposition.

In general, a $2^n$-periodic binary sequence may not have a unique cube decomposition. However numerous examples support the following conjecture.

\noindent {\bf Conjecture  3.1}  Suppose that $s$ is a binary
sequence with period  $2^n$, and $s$ has a standard cube
decomposition without the $t$th order power relation, where $t>1$. If the cubes are in descending
order of   linear complexity, and their dimensions are $m_1,
m_2,\cdots,m_t$, respectively,  then $k$-error linear complexity
will decrease when $k$ is $2^{m_1}$,$2^{m_1}+2^{m_2},\cdots,
2^{m_1}+2^{m_2}+\cdots+2^{m_t}$.

Conjecture  3.1 is of fundamental importance as it provides another
perspective to understand and compute $k$-error linear complexity.

In Example 3.2, though $1+x+x^3+x^4+x^7+x^8$ does not   have a unique cube decomposition,  Conjecture  3.1 still holds.

\

Next we consider the construction of sequences with one or more cubes.
Suppose that $s$ is a binary sequence with period  $2^n$, and
$L(s)=2^n-(2^{i_1}+2^{i_2}+\cdots+2^{i_m})$, where  $0\le i_1<
i_2<\cdots<i_m<n$. We first derive the counting formula of
$m$-cubes with the same linear complexity.

\noindent {\bf Theorem  3.5}  Suppose that $s$ is a binary sequence
with period  $2^n$, and $L(s)=2^n-(2^{i_1}+2^{i_2}+\cdots+2^{i_m})$,
where  $0\le i_1< i_2<\cdots<i_m<n$. If sequence $e$ is an $m$-cube with $L(e)=L(s)$, then the number of  sequence $e$ is
$$2^{2^{m}n-2^{m-1}i_m-\cdots-2i_2-i_1-2^{m+1}+2}$$

\begin{proof}\
Suppose that $s^{(i_1)}$ is a $2^{i_1}$-periodic binary sequence with linear complexity  $2^{i_1}$ and $W_H(s^{(i_1)})=1$,
then  the number of these $s^{(i_1)}$ is $2^{i_1}$

So the number of $2^{i_1+1}$-periodic binary sequences $s^{(i_1+1)}$ with linear complexity $2^{i_1+1}-2^{i_1}=2^{i_1}$ and $W_H(s^{(i_1+1)})=2$ is also $2^{i_1}$.

For $i_2>i_1$,
if $2^{i_2}$-periodic binary sequences $s^{i_2}$ with linear complexity $2^{i_2}-2^{i_1}$ and $W_H(s^{(i_2)})=2$,
then $2^{i_2}-2^{i_1}-(2^{i_1+1}-2^{i_1})=2^{i_2-1}+2^{i_2-2}+\cdots+2^{i_1+1}$.

 Based on Algorithm 2.1,
the number of these $s^{i_2}$ can be given by
$(2^2)^{i_2-i_1-1}\times2^{i_1}=2^{2i_2-i_1-2}$.

For example, suppose that $i_1=1, i_2=3$, then  $(2^2)^{i_2-i_1-1}=4$ sequences

\{1010\ 0000\},
\{1000\ 0010\},
\{010\ 1000\},
\{0000\ 1010\}

of $s^{(i_2)}$
correspond to a sequence \{1010\} of $s^{(i_1+1)}$.

So the number of $2^{i_2+1}$-periodic binary sequences $s^{(i_2+1)}$ with linear complexity $2^{i_2+1}-(2^{i_2}+2^{i_1})=2^{i_2}-2^{i_1}$ and $W_H(s^{(i_2+1)})=4$ is also $2^{2i_2-i_1-2}$.

For $i_3>i_2$, based on Algorithm 2.1,
if $2^{i_3}$-periodic binary sequences $s^{i_3}$ with linear complexity $2^{i_3}-(2^{i_2}+2^{i_1})$ and $W_H(s^{(i_3)})=4$,
then the number of these $s^{i_3}$ can be given by
$(2^4)^{i_3-i_2-1}\times2^{2i_2-i_1-2}=2^{4i_3-2i_2-i_1-2-4}$.

$\cdots\cdots$

So the number of $2^{i_m+1}$-periodic binary sequences $s^{(i_m+1)}$ with linear complexity
$2^{i_m+1}-(2^{i_1}+2^{i_2}+\cdots+2^{i_m})=2^{i_m}-(2^{i_1}+2^{i_2}+\cdots+2^{i_{m-1}})$
and $W_H(s^{(i_m+1)})=2^m$ is also $2^{2^{m-1}i_{m}-\cdots-2i_2-i_1-2-4-\cdots-2^{m-1}}$.

For $n>i_m$,
if $2^{n}$-periodic binary sequences $s^{(n)}$ with linear complexity $2^n-(2^{i_1}+2^{i_2}+\cdots+2^{i_m})$ and $W_H(s^{(n)})=2^m$,
then the number of these $s^{(n)}$ can be given by

\begin{eqnarray*}&&(2^{2^m})^{n-i_m-1}\times2^{2^{m-1}i_{m}-\cdots-2i_2-i_1-2-4-\cdots-2^{m-1}}\\
&=&2^{2^m n-2^{m-1}i_{m}-\cdots-2i_2-i_1-2-4-\cdots-2^{m-1}-2^m}\\
&=&2^{2^m n-2^{m-1}i_{m}-\cdots-2i_2-i_1-2^{m+1}+2}
\end{eqnarray*}

\end{proof}\

For $2^n$-periodic binary sequences $s$ and $e$, if
$W_H(e)=k_{\min}$ and $L(s+e)<L(s)$, then  the sequence $e$ is called
as a $k$-error vector. By  cube theory, a
$k$-error vector is in fact an $m$-cube with the same linear
complexity $L(s)$.

\

Etzion et al.
proved Theorem 3 in \cite{Etzion}, which is  equivalent to Theorem 3.5, with a much different approach. The approach here is much simpler.

Suppose that $s$ is a $2^n$-periodic binary sequence
with more than one cube,  and  each cube has a fixed linear complexity.
Now we consider  the counting formula of these sequences.

\noindent {\bf Theorem   3.6} Suppose that $s$ is a $2^n$-periodic binary sequence
with two independent cubes: $C_1,C_2$.  $C_1$ has linear complexity
 $2^n-(2^{i_1}+2^{i_2}+\cdots+2^{i_m})$,
where  $0\le i_1< i_2<\cdots<i_m<n$, and  $C_2$ has linear complexity
$2^n-(2^{j_1}+2^{j_2}+\cdots+2^{j_l})$, where  $0\le j_1< j_2<\cdots<j_l<n$ and $2^{j_1}>2^{t}$, where $t=\max\{x\mid i_x\le j_1,  x\ge1\}$.
Then the number of sequence $s$ is {
\small
$(2^{2^{m}n-2^{m-1}i_m-\cdots-2i_2-i_1-2^{m+1}+2})[2^{2^{l}n-2^{l-1}j_l-\cdots-2j_2-2j_1-2^{l+1}+2}$ $(2^{j_1}-2^{t})]$}

\begin{proof}\
The proof is very similar to that of Theorem   3.5.

Noted that  $s^{(j_1)}$ is a $2^{j_1}$-periodic binary sequence with linear complexity  $2^{j_1}-(2^{i_1}+2^{i_2}+\cdots+2^{i_{t}})$ and $W_H(s^{(j_1)})=2^{t}$,
  the number of  zero elements  in $s^{(j_1)}$  is $2^{j_1}-2^{t}$.

Similar to the proof of Theorem   3.5, for each cube $C_1$, the number of cube $C_2$ is $2^{2^{l}n-2^{l-1}j_l-\cdots-2j_2-2j_1-2^{l+1}+2}$ $(2^{j_1}-2^{t})$

This completes proof.

\end{proof}\

Next we give an example to illustrate Theorem 3.6.

\noindent {\bf Example 3.4 } Suppose that $s$ is a $2^3$-periodic binary sequence
with 2 independent cubes: $C_1,C_2$.  $C_1$ has linear complexity
 $2^n-2^{0}$,
 and  $C_2$ has linear complexity
$2^n-2^{2}$, where   $2^{2}>2^{1}$ and $t=1$. From sequence 11, we get 1100,0110,1001,0011, and from sequence 1100, we get 11000000,01001000,10000100,00001100. For each cube $C_1$, the number of cube $C_2$ is 2.  For 11000000, we get 11100010 and 11010001. The total number is 32. The result is consistent with Theorem 3.6.

It should be noted that the main idea in the proof of Theorem 3.6 is that two cubes are  relatively independent. So after determining the connecting part of two cubes, one can construct each cube independently.
Given the linear complexities of $t(t>2)$ cubes,  with the approach of Theorem   3.6, we can discuss the  number of sequence $s$ with $t$ cubes, of which each cube has a fixed linear complexity. Here we only give the results for $t=3$.

\noindent {\bf Corollary   3.3} Suppose that $s$ is a $2^n$-periodic binary sequence
with 3 independent cubes: $C_1,C_2$ and $C_3$.  $C_1$ has linear complexity
 $2^n-(2^{i_1}+2^{i_2}+\cdots+2^{i_m})$,
where  $0\le i_1< i_2<\cdots<i_m<n$,   $C_2$ has linear complexity
$2^n-(2^{j_1}+2^{j_2}+\cdots+2^{j_l})$, where  $0\le j_1< j_2<\cdots<j_l<n$ and $ 2^{j_1}>2^{t}$, where $t=\max\{x\mid i_x\le j_1,  x\ge1\}$.
and
  $C_3$ has linear complexity
$2^n-(2^{k_1}+2^{k_2}+\cdots+2^{k_w})$, where  $0\le k_1< k_2<\cdots<k_w<n$ and $2^{k_1}>2^{u}+2^{v}$, where $u=\max\{x\mid i_x\le k_1,  x\ge1\}$ and $v=\max\{y\mid j_y\le k_1,  y\ge1\}$.
Then the number of sequence $s$ is {
\small
$(2^{2^{m}n-2^{m-1}i_m-\cdots-2i_2-i_1-2^{m+1}+2})[2^{2^{l}n-2^{l-1}j_l-\cdots-2j_2-2j_1-2^{l+1}+2}$ $(2^{j_1}-2^{t})]
[2^{2^{w}n-2^{w-1}k_w-\cdots-2k_2-2k_1-2^{w+1}+2}$ $(2^{k_1}-2^{u}-2^{v})]$}

\begin{proof}\
The proof is very similar to that of Theorem   3.6.

Noted that  $s^{(j_1)}$ is a $2^{j_1}$-periodic binary sequence with linear complexity  $2^{j_1}-(2^{i_1}+2^{i_2}+\cdots+2^{i_{t}})$ and $W_H(s^{(j_1)})=2^{t}$,
  the number of  zero elements  in $s^{(j_1)}$  is $2^{j_1}-2^{t}$,
 and
  $s^{(k_1)}$ is a $2^{k_1}$-periodic binary sequence with $W_H(s^{(k_1)})=2^{u}+2^{v}$,
  the number of  zero elements  in $s^{(k_1)}$  is $2^{k_1}-2^{u}-2^{v}$.

Similar to the proof of Theorem   3.5, for each cube $C_1$ and $C_2$, the number of cube $C_3$ is $2^{2^{w}n-2^{w-1}k_w-\cdots-2k_2-2k_1-2^{w+1}+2}$ $(2^{k_1}-2^{u}-2^{v})$

This completes proof.

\end{proof}\

It should be noted that the mild requirement $2^{j_1}>2^{t}$ in
Theorem 3.6 is not critical and the idea of constructing cubes independently can
also be used for cases not meeting the condition,
one can use the similar approach to find the cube number. We
illustrate these cases by the following example.

\noindent {\bf Example 3.5 } We now  construct
$2^n$-periodic binary sequence
with 2 independent cubes: $C_1,C_2$.  $C_1$ has linear complexity $2^n-(1+2)$
 and  $C_2$ has linear complexity
$2^n-(1+8)$. The number of $2^2$-periodic binary sequence with linear complexity $2^n-(1+2)$ is 1. Namely sequence 1111.
The number of $2^3$-periodic binary sequence with linear complexity $2^n-(1+2)$ is $2^4$. For instance sequence 11110000.
Each $2^3$-periodic binary sequence has 4 options to put 2 non-zero elements with distance 1.  The number of $2^4$-periodic binary sequence
with 2 independent cubes is $2^4\times 2^4 \times 4$.

Finally the number of $2^n$-periodic binary sequence
with 2 independent cubes is $2^4\times 2^4 \times 4 \times (2^8)^{n-4}=2^{10}\times (2^8)^{n-4}$.

For $n=4$, one example is 1111 1100 1100 0000.

\section{Conclusion}

A small number of element changes may lead to a sharp decline of
linear complexity, so the concept of stable $k$-error linear
complexity has been introduced in this paper. By studying the linear
complexity of binary sequences with period $2^n$, especially the
linear complexity may decline when the superposition of two
sequences with the same linear complexity,  a new approach to construct the sequence with stable
$k$-error linear complexity based on cube theory has been derived.
It has been proved that a binary sequence with period  $2^n$ can
be decomposed into several disjoint cubes and further  a standard cube decomposition approach has been presented, so a new approach to
study $k$-error linear complexity has been introduced.

Etzion et al. \cite{Etzion}  studied
sequences  only having two $k$-error linear complexity
  values exactly, either its $k$-error linear complexity is
only $L(s)$ or 0. So these sequences possess stable $k$-error linear
complexity, but not necessarily the maximum stable $k$-error linear
complexity. We extended the results reported in \cite{Etzion}.

From the perspective of cube theory, one can easily perceive the  real problem and difficulty points of the  $k$-error linear
complexity for a binary sequence with more than one cube.

In future,  we may further investigate the essential relationship between the standard cube decomposition and  the  $k$-error linear
complexity of a binary sequence with period $2^n$ and try to prove
Conjecture 3.1.

 \section*{ Acknowledgment}
 The research was supported by
Anhui Natural Science Foundation(No.1208085MF106) and NSAF
(No. 10776077).

\end{document}